\documentclass[12pt,preprint]{aastex}

\shorttitle{Images of Vega Dust Ring}
\shortauthors{Marsh et al.}

\begin{document}

\title{Images of Vega Dust Ring at 350 and 450 $\mu$m:  New Clues to the
Trapping of Multiple-Sized Dust Particles in Planetary Resonances}

\author{K. A. Marsh\altaffilmark{1}, 
C. D. Dowell\altaffilmark{1}, 
T. Velusamy\altaffilmark{1}, 
K. Grogan\altaffilmark{1},
C. A. Beichman\altaffilmark{2}}

\altaffiltext{1}{Jet Propulsion Laboratory, 4800 Oak Grove Drive, Pasadena,
CA 91109; Kenneth.A.Marsh@jpl.nasa.gov, cdd@submm.caltech.edu, 
Thangasamy.Velusamy@jpl.nasa.gov, Keith.Grogan@jpl.nasa.gov}
\altaffiltext{2}{California Institute of Technology, IPAC, MS 100-22,
Pasadena, CA 91125; chas@ipac.caltech.edu}

\pagebreak

\begin{abstract}
We have used the SHARC II camera at Caltech Submillimeter Observatory
to make 350 $\mu$m and 450 $\mu$m images of the Vega dust disk at spatial
resolutions (FWHM) of 9\farcs7 and 11\farcs1, respectively.  The images show
a ring-like morphology (radius $\sim100$ AU) with inhomogeneous structure 
that is qualitatively different from that previously reported at 850 $\mu$m 
and longer wavelengths.  We attribute the 350/450 $\mu$m emission to
a grain population whose characteristic size ($\sim1$ mm) is intermediate
between that of the cm-sized grains responsible for emission longward of
850 $\mu$m and the much smaller grains ($\stackrel{<}{_\sim}18$ $\mu$m) in
the extensive halo, visible at 70 $\mu$m, discussed by \citet{su05}.
We have combined our submillimeter images with 
{\em Spitzer\/} data at 70 $\mu$m to produce 2-d maps of line-of-sight 
optical depth (relative column density). These ``tau maps" suggest that the 
mm-sized grains are located preferentially in three symmetrically-located
concentrations.  If so, then this structure could be understood in
terms of the \cite{wya03} model in which planetesimals are trapped in the 
mean motion resonances of a Neptune-mass planet at 65 AU, provided
allowance is made for the spatial distribution of dust grains to differ from 
that of the parent planetesimals. The peaks of the tau maps
are, in fact, located near the expected positions corresponding to 
the 4:3 resonance.  If this
identification is confirmed by future observations, it would resolve an 
ambiguity with regard to the location of the planet.
\end{abstract}


\section{Introduction}

The observable behavior of circumstellar dust particles under the influence of 
gravity and radiation pressure can provide information on
the locations and masses of unseen planets (see, for example,
\citet{wya03,wya06,del05}). A particularly suitable object for study
is the A0 star Vega at 7.76 pc whose disk is seen nearly face-on, as suggested
by the low inclination ($5^\circ$) of the stellar rotation axis \citep{gul94}.
Images at 850 $\mu$m and 1.3 mm 
\citep{hol98,wil02,koe01} indicate a partial dust ring ($r\sim100$ AU) 
dominated by two unequal clumps
interpreted as collisional debris from resonantly-trapped 
planetesimals. More recent observations with the Multiband Imaging
Photometer for {\em Spitzer\/} (MIPS) \citep{su05} have revealed an
extensive halo representing small ($\sim2$ $\mu$m) and medium-sized 
($\sim18$ $\mu$m) grains blown out from the ring by radiation pressure, the 
ring itself being dominated by large grains of radii $a>180$ $\mu$m.

Dynamical modeling shows the spatial distribution of the ring to be consistent 
with the 2:1(u) resonance
of a Neptune-mass planet which has undergone migration \citep{wya03}. 
Such a model predicts the existence of other populated resonances (for example 
3:2 and 4:3), the detection of which would help confirm the model and
provide new constraints.  We present new observations
at 350 $\mu$m and 450 $\mu$m which bear on this issue.

\section{Observations and Data Reduction}

Vega was observed at 350 $\mu$m and 450 $\mu$m with the SHARC II
camera \citep{dow03} at the Nasmyth focus of the Caltech Submillimeter
Observatory (CSO) on UT 2004 Sep 17--19 (350 $\mu$m; $\tau_{\rm 225\,GHz}
\simeq 0.041$), 2005 Apr 23 (350 $\mu$m; $\tau_{\rm 225\,GHz}\simeq 0.036$),
and 2005 Jun 11--13 (450 $\mu$m; $\tau_{\rm 225\,GHz}\simeq 0.048$), where
$\tau_{\rm 225\,GHz}$ represents the zenith value of atmospheric optical
depth.  The on-source total integration times were 5.0 hr, 2.7 hr and
9.7 hr, respectively.  Similarly to our Fomalhaut observations
\citep{mar05}, the telescope was scanned in an oscillatory fashion in azimuth
and elevation with peak-to-peak amplitudes of $30''-100''$. The Vega images
were generated with an iterative code (``sharcsolve") similar to 
CRUSH\footnote{See http://www.submm.caltech.edu/~sharc/crush/index.htm}. 
To reduce spatial $1/f$ noise in the center of the image, the sky intensity 
was forced to be zero beyond $35''$ from the star (analogous to spatial 
chopping with amplitude $70''$ in all directions), except for the final
iteration which allowed non-zero intensities outside $35''$ radius for 
noise-evaluation purposes.  After the application of 
Gaussian smoothing (with a kernel of 5\farcs5 FWHM),
the angular resolutions (FWHM) of the output images were 
9\farcs7 and 11\farcs1, at 350 $\mu$m and 450 $\mu$m, respectively.  
Absolute calibration was accomplished with hourly, interspersed observations 
of point sources, and is based on assumed Neptune fluxes of $S_{350}=92.2$ and
86.5 Jy in 2004 and 2005, respectively, and 
$S_{450}=65$ Jy.  We estimate 1$\sigma$ calibration accuracies of 30\% and
$1''$ in absolute flux and pointing, respectively. The
flux error includes the effect of subtraction of the slowly-varying background
($\sim5$\% level over spatial scales of a few tens of arcseconds at both
wavelengths).

The point source response function (PSF) was obtained from observations of 
Mars at 350 $\mu$m and Neptune at 450 $\mu$m, rotated and coadded to recreate 
the rotational smearing due to changing parallactic angle. 
The nonzero angular diameters of those planets
(3\farcs5 and 2\farcs3, respectively) broadened
the estimated PSFs by 6\% and 2\%, respectively; the effect
on subsequent processing was found to be negligible within prevailing
errors.

Figure \ref{fig1} shows the observed images at 350 $\mu$m
and 450 $\mu$m, before and after subtracting the estimated 
photospheric contributions of 35 mJy and 21 mJy, respectively.
The integrated flux densities of Vega (before photospheric subtraction), 
within a circular aperture of $30''$
radius, at 350 $\mu$m and 450 $\mu$m are $500\pm150$ mJy and $150\pm45$ mJy,
respectively. 

The observed images show clearly the ring morphology of the 
Vega disk, and suggest inhomogeneous structure.  Figure 2
shows the 350 $\mu$m intensity variation as a function of azimuth,
calculated in an annulus of inner and outer radii 6\farcs9 and 13\farcs9,
respectively.  Three peaks are apparent, at azimuths of $-90^\circ$, $0^\circ$,
and $120^\circ$, with amplitudes of $3\sigma$, $4\sigma$, and $2\sigma$
above the local mean level respectively, where $\sigma$ represents the
statistical measurement noise.  

\section{Mapping the Relative Column Density of Dust}

We have mapped the line-of-sight optical depth (relative column density of 
dust) in the Vega disk as a function of 2-d location in the disk plane using
the DISKFIT procedure \citep{mar05}, assuming in this case
a geometrically-thin face-on disk.  The output, referred to as
a ``tau map," is estimated using a set of observed images at 
multiple wavelengths, taking full account of the corresponding PSFs. 
It is assumed that the
local temperature of each dust grain component is determined by the
energy balance of individual grains in the stellar radiation field using
the results of \citet{bac93}.  The current version of the code makes
simultaneous estimates of the tau maps corresponding to the different
grain components, each of which is characterized by the parameters $\lambda_0$
and $\beta$, where $\lambda_0$ represents the wavelength above which the
grains radiate inefficiently and has an approximate correspondence with the
grain radius, $a$, and $\beta$ is the power-law index of the wavelength 
dependence of opacity such that $\kappa_\lambda\propto\lambda^{-\beta}$. 
The spatial resolution of the tau maps exceeds that of 
the raw images due to implicit deconvolution of the PSFs.

Tau maps were made using data at three wavelengths, by combining our 
350 \& 450 $\mu$m images with a 70 $\mu$m MIPS image at the fine ($5''$)
pixel scale. We employed a two-grain model consisting
of the 18 $\mu$m grains which dominate the 70 $\mu$m emission
\citep{su05} and a population of larger grains whose size was chosen based 
primarily on the observed spectral slope between 350 $\mu$m and 70 $\mu$m. 
Specifically, assuming $\beta=1$ \citep{den00}, the spectral slope 
implies $T_{\rm dust}\stackrel{<}{_\sim} 50$ K for the larger grains;
if we further assume a size-temperature dependence corresponding to the grain
composition (silicate-carbon mix) used by \citet{su05} in their Vega modeling, 
we obtain $a\stackrel{>}{_\sim}100$ $\mu$m.  Since this
limit was based on the extremes of the flux error bars and of the relative
contributions of ring and halo at 70 $\mu$m, we have
adopted a larger value (1 mm) as being more likely to be representative of
the grains reponsible for 350/450 $\mu$m emission. The results
are presented in Figure \ref{fig3}, which shows tau maps 
for the two grain components (18 $\mu$m and 1 mm)
separately and also the total optical depth (lower panel). 
Superposed on the latter, for comparison, are the positions of 
previously-reported emission peaks from longer wavelength data at 850 $\mu$m 
and 1.3 mm \citep{hol98,koe01,wil02}. 

The tau map for 1 mm grains shows inhomogeneities, the reality of which
we have assessed using a $\chi^2$ test based on fits to the 350 $\mu$m
and 450 $\mu$m data. Specifically, an azimuthally-uniform ring resulted in
a 10\% increase in $\chi^2$ relative to an unconstrained tau map; taking into 
account the number of independent data points (255), this translates into
a relative probability $\sim5\times10^{-6}$ corresponding to a $4.6\sigma$
deviation.  We therefore conclude that the azimuthal structure we see in the
ring is at the 4$\sigma$--5$\sigma$ significance level, consistent with our 
findings based on the $S/N$ of peaks in the observed 350 $\mu$m image 
(see Figure \ref{fig2}).

The tau map for 18 $\mu$m grains is dominated
by a ring of larger scale ($r\simeq140$ AU) than for
the 1 mm grains ($r\simeq100$ AU).  There is also lower-level diffuse 
structure masked by surrounding noise spikes, the latter resulting from
the radial increase in tau estimation error corresponding to decreasing dust 
temperature.  We can, however, 
smooth out these fluctuations with azimuthal averaging to
show the underlying diffuse structure.  Figure \ref{fig4} shows the result
in the form of radial profile plots of
the 1 mm grain component (filled circles) and the 18 $\mu$m grain
component (open circles).  The error bars reflect the tau map estimation
errors after a $\sqrt{N}$ reduction from azimuthal averaging in
annuli of width 17.5 AU. It is apparent that for
$r>200$ AU, the radial falloff in column density of 18 $\mu$m grains 
is consistent with the $1/r$ variation reported by \citet{su05}. Inside
200 AU, however, some of these grains may be concentrated in the
the ring structure, but the proportion is subject to model uncertainties 
related to the larger grain component.

\section{Interpretation}

The observations show qualitatively different appearance of 
the Vega dust ring at 350/450 $\mu$m and $\ge850$ $\mu$m. Furthermore,
the 850 $\mu$m flux (23 mJy) predicted by our tau map accounts for only a
fraction of either of the published values of 850 $\mu$m flux, which 
are $45.7\pm5.4$ Jy \citep{hol98} and 91 mJy (\citet{hol05}, as quoted
by \citet{su05}).  The observations can, however, be reconciled using a
model in which the ring emission at 350/450 $\mu$m is dominated by
1 mm grains with a $\lambda^{-1}$ opacity law, and the emission longward
of 850 $\mu$m is dominated by an additional component of much larger grains 
with flat spectral opacity ($\beta\sim0$); the latter would require
grain sizes of at least a centimeter \citep{pol94}.

The morphology of the 
850 $\mu$m image (with its single dominant peak at $\sim45^\circ$ in position 
angle) has been attributed to dynamical effects, the most important of which is
the 2:1(u) resonance of a Neptune-mass planet 65 AU from the star \citep{wya03}.
Those calculations also show that other resonances are present, principally
3:2 and 4:3,  the latter of which shows a 3-fold spatial symmetry
consistent with what we see at 350/450 $\mu$m.  In this regard we note that
for a planet of given mass, the requirement that the dust grains not be
dislodged from resonance by radiation pressure sets a lower limit, 
$a_{\rm crit}$, to the grain size \citep{wya06}.  For a Neptune-mass planet
orbiting Vega (luminosity $60L_\odot$; mass $2.5M_\odot$),
Wyatt's expression leads to $a_{\rm crit}=900$ $\mu$m,
consistent with our choice of $a=1$ mm for the tau maps.

Based on the 850 $\mu$m data alone, an ambiguity exists for the orbital 
direction of the planet, shown counterclockwise in Figure 15(b) of 
\citet{wya03}.  But if our identification of the 
350/450 $\mu$m structure with the 4:3 resonance is correct, then 
the ambiguity is resolved; the planet is then located diametrically opposite to
the star from the SE clump in our Figure \ref{fig3} and its orbit is
clockwise. The predicted locations
of the 4:3 resonance peaks are then given by the arrows in Figure \ref{fig2},
which are in substantial agreement with observation.

It would then remain to explain why the cm-sized grains (seen at 850 $\mu$m) and
mm-sized grains (seen at 350/450 $\mu$m) exhibit two different resonance
patterns.  We note, however, that the analysis by \citet{su05} suggests that
the dust has been released during a relatively recent event, most likely
the collision of two planetesimals.  This being the case, our results could
be understood if the two planetesimals involved were located in the 4:3 
resonance, and that the collision released smaller
particles than the ones which were there before, thus affecting the
350/450 $\mu$m appearance much more than that at 850 $\mu$m.
The 4:3 resonance is, in fact, a relatively likely place for a collision to 
occur, since the collision timescale decreases with decreasing orbital 
semimajor axis, and hence is shorter in the 4:3 resonance than in 2:1.

\begin{figure}
\epsscale{0.95}
\plotone{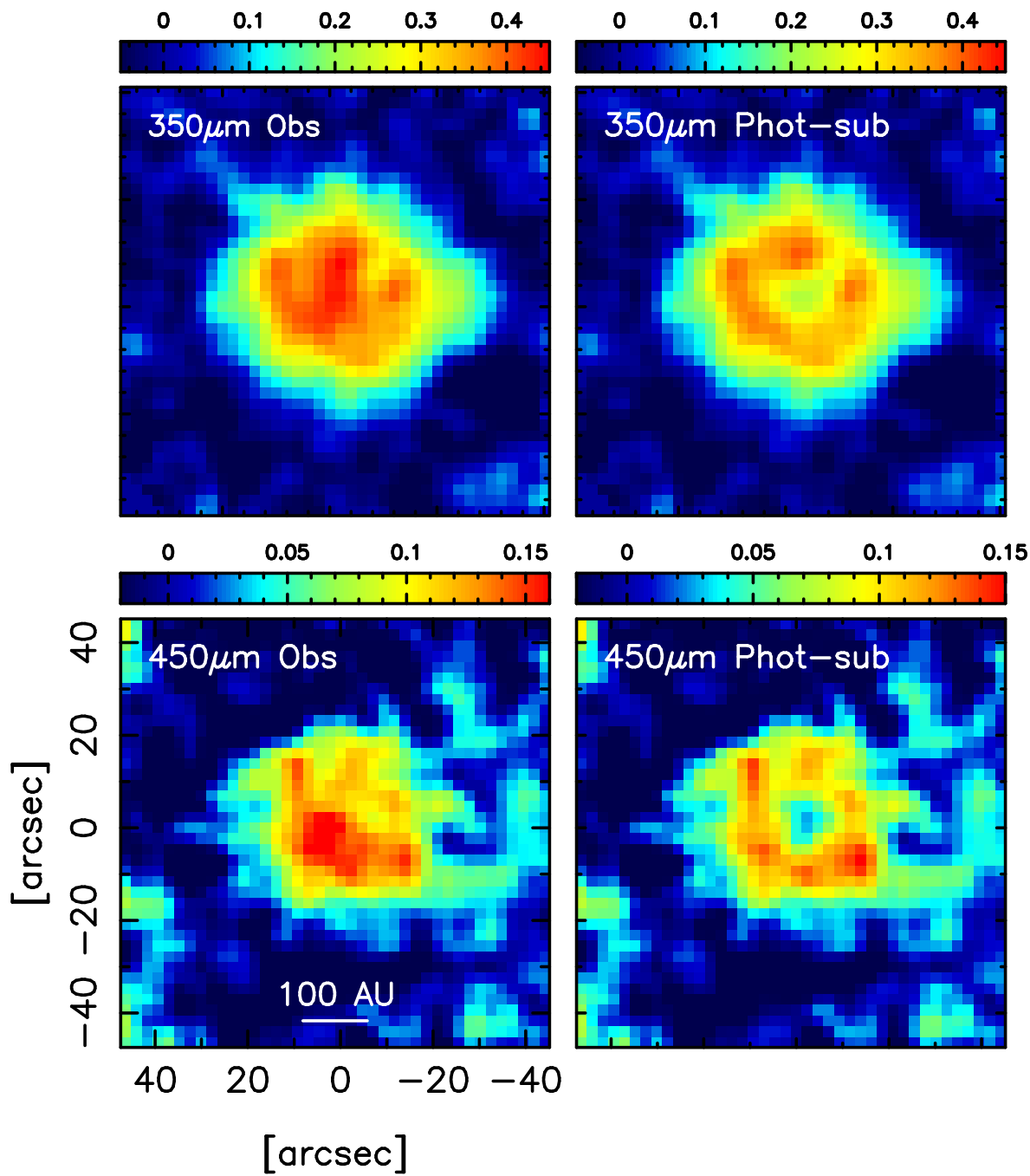}
\caption{Observed and photosphere-subtracted images of Vega at 350 $\mu$m 
and 450 $\mu$m. The intensity scale (shown by the horizontal bar at the top
of each image) is in units of mJy arcsec$^{-2}$ and the orientation is
such that north is up and east is to the left.  The RMS measurement noise on 
the images is 0.032 and 0.026 mJy arcsec$^{-2}$ at 350 $\mu$m and 450 $\mu$m, 
respectively.
\label{fig1}}
\end{figure}

\begin{figure}
\plotone{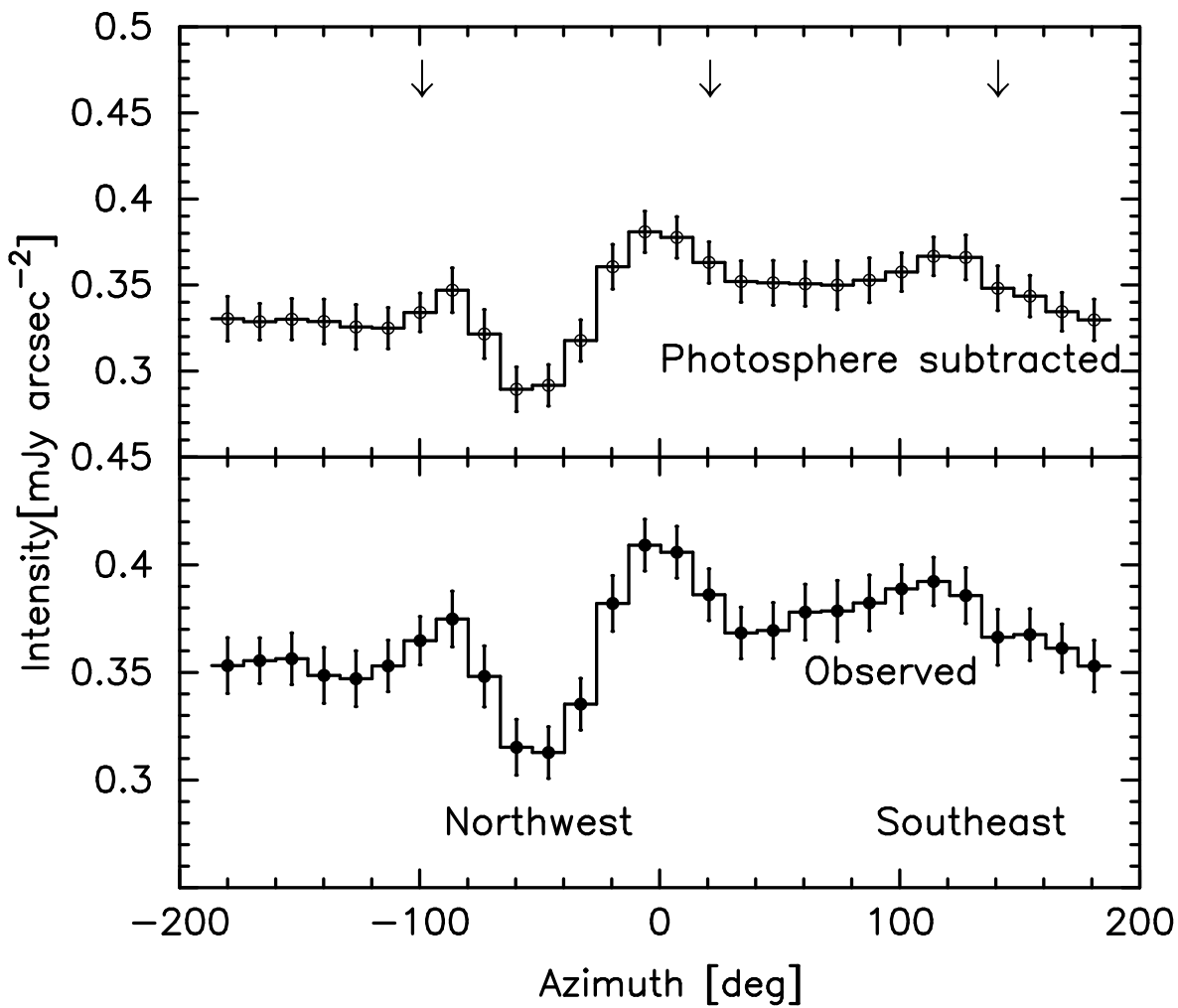}
\caption{Intensity in the Vega disk at 350 $\mu$m wavelength as a function of
azimuth around the ring (measured east from north), both with and without 
subtraction of the estimated photospheric contribution.  Error bars represent 
the standard deviation of measurement noise.  The arrows represent the
expected locations of the 4:3 resonance peaks for the dynamical model of
\cite{wya03} in which dust is trapped in the mean motion resonances of a 
suspected planet. \label{fig2}}
\end{figure}

\begin{figure}
\epsscale{0.44}
\plotone{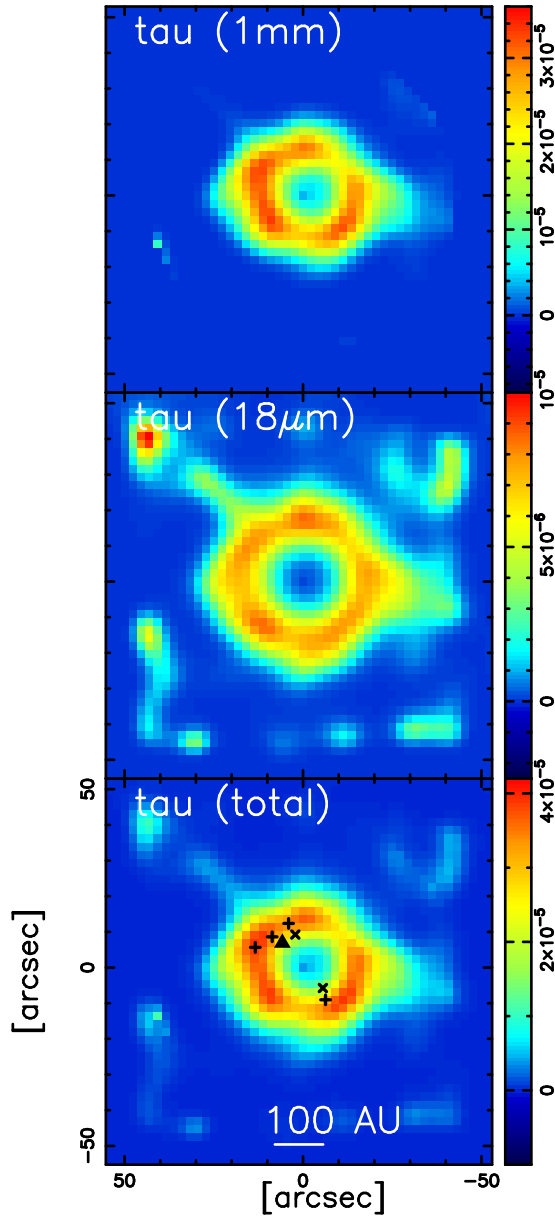}
\caption{The estimated line-of-sight optical depth of the Vega disk at a 
reference wavelength of 350 $\mu$m, based on observed images at 350 $\mu$m, 
450 $\mu$m, and 70 $\mu$m, assuming a two-component dust grain model 
in which the grain radii are 1 mm and 18 $\mu$m. The upper two panels show the
optical depth distribution for the two grain components separately, while
the lower panel shows the sum of both components.  Also shown on the 
lower plot are the locations of previously reported emission peaks 
from longer wavelength data as follows: (filled triangle)
850 $\mu$m---\citet{hol98}; ($\times$) 1.3 mm---\citet{wil02}; ($+$) 
1.3 mm---\citet{koe01}. \label{fig3}}
\end{figure}

\begin{figure}
\epsscale{1.0}
\plotone{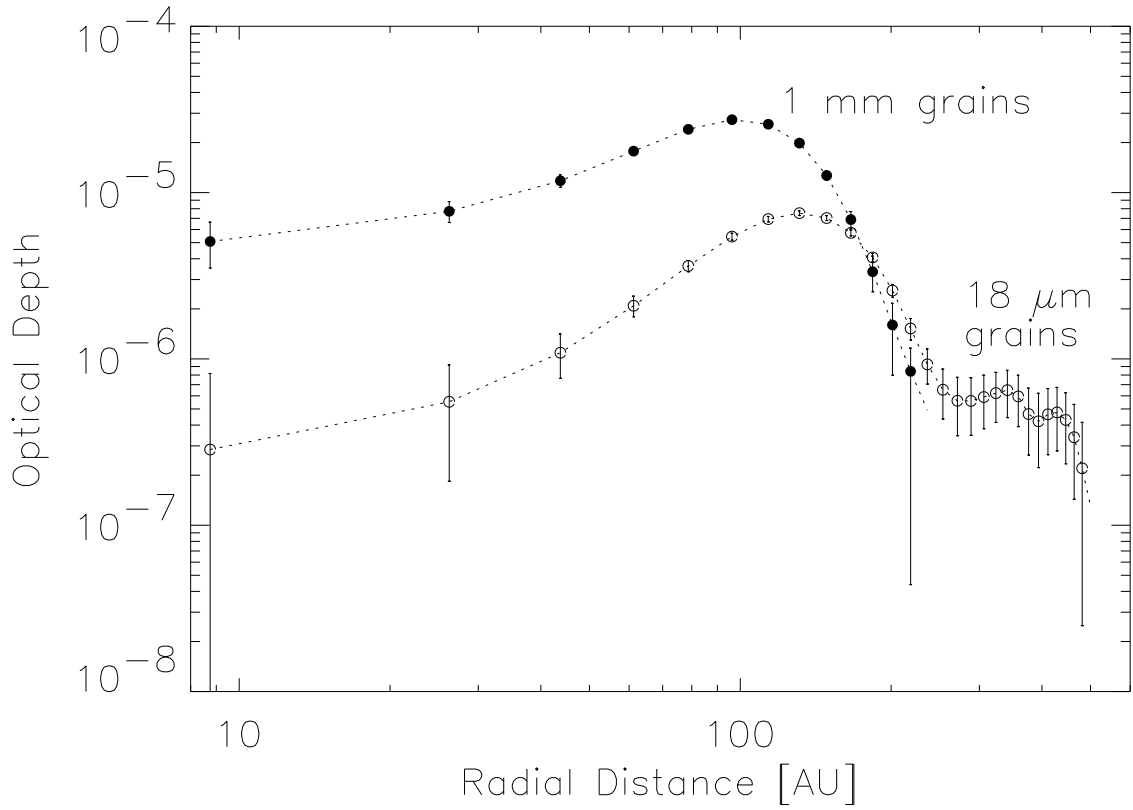}
\caption{Radial profiles of estimated optical depth for the two grain 
components, characterized by grain radii of 1 mm (filled circles) and 
18 $\mu$m (open circles). 
\label{fig4}}
\end{figure}

\begin{figure}
\epsscale{0.69}
\plotone{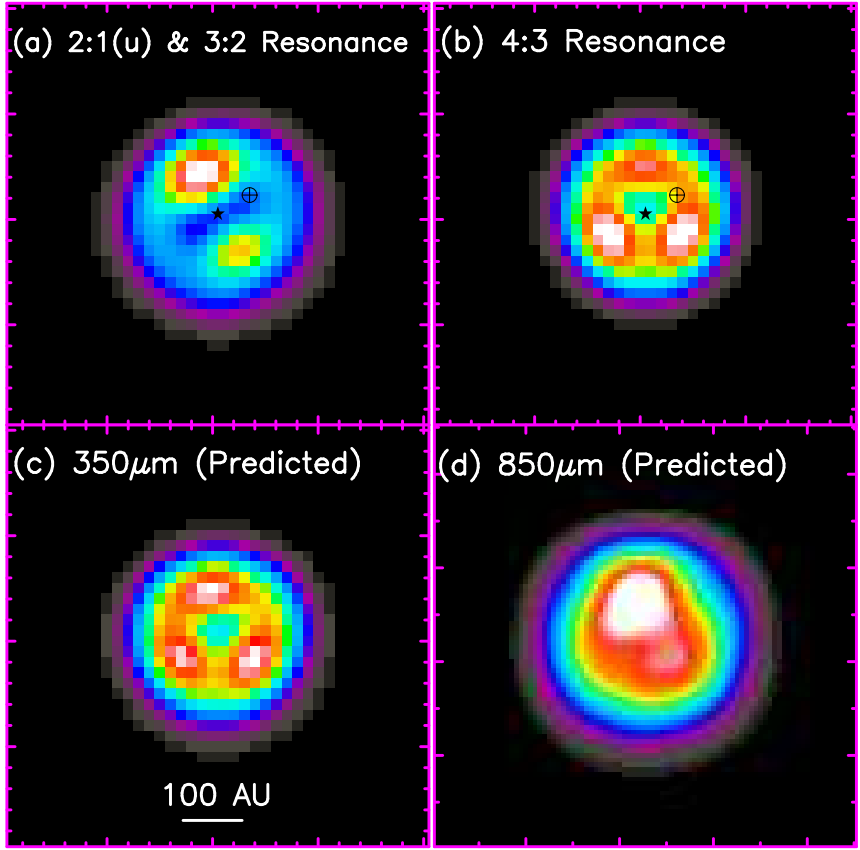}
\caption{Results of dynamical modeling of the Vega dust disk showing the
predicted intensity distributions of dust emission at 350 $\mu$m and 850 $\mu$m.
Following \cite{wya03} we assumed a Neptune-mass planet at 65 AU, and
further assumed that the planet orbits 
clockwise, is currently located NW of the star and that its 4:3 resonance is 
populated by smaller grains (1 mm) than the other resonances (1 cm).  
{\em Upper two panels:\/} The 350 $\mu$m emission resulting from (a) the 
combined effects of the 2:1(u) and 3:2 resonances, and (b) the 
4:3 resonance.  The planet is indicated by an encircled cross.
{\em Lower two panels:\/} Predicted observational images at (c) 350 $\mu$m and
(d) 850 $\mu$m. Images (c) and (d) were produced by adjusting the 
relative amounts of material in the two grain populations (1 mm and 1 cm)
to provide the best match to the observed images; this required
equal amounts of material (by cross-sectional area) in each.
In this model, the 1 mm grains (4:3 resonance) contribute $80\pm16$\% of 
the observed flux at 350 $\mu$m and $60\pm12$\% at 850 $\mu$m, 
while the 1 cm grains (2:1(u) \& 3:2 resonances) contribute 
20\% at 350 $\mu$m and 40\% at 850 $\mu$m. 
All of the above images have been
smoothed to the spatial resolution of the observations,
corresponding to FWHM of $10''$ at 350 $\mu$m (CSO/SHARC II; present paper) and
$14''$ at 850 $\mu$m (JCMT/SCUBA; \citet{hol98}).
On the linear pseudocolor intensity scale, white corresponds to peak intensity 
for each individual image.  
\label{fig5}}
\end{figure}

We have simulated this scenario with dynamical modeling based on a
Neptune-mass planet at 65 AU, assuming orbital parameters similar to 
\citet{wya03}, and further assuming that the 4:3 resonance is populated by 
smaller grains (1 mm) than the other resonances (1 cm).  The results are 
presented in Figure \ref{fig5}. Comparison of the predicted 350 $\mu$m
and 850 $\mu$m images with the observations
(Figure \ref{fig1} of this paper and \citet{hol98}, respectively)
shows that the model successfully accounts for the wavelength dependence
of the observed structure.
Our analysis therefore suggests that
different resonances can be populated by grain populations of different sizes,
depending on the collisional history of the planetesimals. This means that
the dust distribution does not necessarily mimic the planetesimal
distribution as assumed by \citet{wya03}. It will thus be important to confirm 
the detection of the 4:3 resonance and constrain the grain sizes involved
using future observations at multiple wavelengths.

\acknowledgments

This work was performed by the Jet Propulsion Laboratory, 
California Institute of Technology, under contract with the 
National Aeronautics and Space Administration.  Research at the Caltech
Submillimeter Observatory is supported by NSF grant AST-0229008.

\clearpage

\clearpage

\clearpage

\clearpage

\clearpage

\end{document}